\documentclass[aps,prl,twocolumn, amsmath,amssymb,,showpacs,superscriptaddress,groupedaddress,floatfix, notitlepage]{revtex4-1} 

\usepackage{graphicx}
\usepackage{dcolumn}
\usepackage{bm}
\usepackage{xcolor}

\begin{document}


\title{Hyperuniformity in cyclically driven glasses}

\author{Saheli Mitra}\affiliation{Universit\'e Paris-Saclay, CNRS, Laboratoire de Physique des Solides, 91405 Orsay, France.}%
\author{Anshul D. S. Parmar}
\thanks{Present address: \textit{Laboratoire Charles Coulomb (L2C), Universit\'e de Montpellier, CNRS, 34095 Montpellier, France.}}
\affiliation{\textit{Theoretical Sciences Unit, Jawaharlal Nehru Centre for Advanced Scientific Research, Jakkur Campus, Bengaluru 560 064, India.}}
\author{Premkumar Leishangthem}
\thanks{Present address: \textit{Complex Systems Division, Beijing Computational Science Research Center, Beijing 100193, China}}
\affiliation{\textit{Theoretical Sciences Unit, Jawaharlal Nehru Centre for Advanced Scientific Research, Jakkur Campus, Bengaluru 560 064, India.}}
\author{Srikanth Sastry}%
\email{sastry@jncasr.ac.in}
\affiliation{\textit{Theoretical Sciences Unit, Jawaharlal Nehru Centre for Advanced Scientific Research, Jakkur Campus, Bengaluru 560 064, India.}}

\author{Giuseppe Foffi}
\email{giuseppe.foffi@u-psud.fr}
\affiliation{Universit\'e Paris-Saclay, CNRS, Laboratoire de Physique des Solides, 91405 Orsay, France.}%
\newcommand{\GF}[1]{\textcolor{blue}{#1}}
\newcommand{\ADSP}[1]{\textcolor{cyan}{#1}}
\date{\today}

\begin{abstract}

We present a numerical investigation of the density fluctuations in a model glass under cyclic shear deformation.  At low amplitude of shear, below yielding, the system reaches a steady absorbing state in which density fluctuations are suppressed revealing a clear fingerprint of  hyperuniformity up to a finite length scale. The opposite scenario is observed above yielding, where the density fluctuations are strongly enhanced. {We demonstrate that the transition to  this  state is accompanied by a spatial phase separation into  two {distinct} hyperuniform {regions}, as a consequence of shear band formation above the yield amplitude.}

\end{abstract}

\maketitle

The last decade has seen a  growing  interest in the  idea of `hyperuniformity'  in several domains of physics~\cite{Torquato2018}. This general concept is related to  the suppression of large length scale density fluctuations  in  disordered systems.  First introduced in the context of  matter distribution in early universe~\cite{gabrielli2002}, the concept of hyperuniformity was later extended to disordered condensed matter physics~\cite{torquato2003local}. The  idea is to consider how a set of points, embedded in a space of dimensionality $d$, is distributed in a volume of varying size, for example a sphere of radius $R$. For a  completely random point distribution, the variance of number density  within  the observation window scales as $R^{-d}$. For points on a regular lattice, however,  the variance scales as $R^{-(d+1)}$. A hyperuniform disordered system will represent an intermediate case between these two, with a variance scaling as  $R^{-(d+\alpha)}$ with $\alpha$ having positive values.  This peculiar behavior is directly related to a vanishing structure factor in the long wavelength  limit, i.e.   $S(k\rightarrow 0) \rightarrow 0$  and, as  a consequence, a vanishing isothermal compressibility $\chi_T$~\cite{torquato2003local}. 
\\
On account of this peculiar behavior, hyperuniform disordered systems are considered as a new exotic state of matter \cite{Torquato2018}. This seems to be justified by the fact that, since its introduction, the concept of  hyperuniformity has been discussed in several contexts such as  jammed  packings~\cite{Donev2005, berthier2011suppressed},  biological tissues~\cite{jiao2014avian, Klatt2019}, superconductors \cite{LeThien2017} and colloids~\cite{Kurita2011}.  A strong  interest in hyperuniform systems come also from their photonic properties as they display complete photonic band gaps \cite{Florescu2009,  FroufePrez2016, Sellers2017}, stealth material\cite{torquato2015ensemble} and their use as effective wave guides~\cite{Leseur2016}. 
\\
In recent years, hyperuniform behavior has been also found in several driven systems undergoing non-equilibrium transitions from a diffusive state to an absorbing state. In this  situation the absorbing states have been found to be hyperuniform in the context of emulsions~\cite{weijs2015emergent}, suspensions~\cite{tjhung2015hyperuniform,Hexner2017prl,hexner2017,wang2018} and active particle systems~\cite{Lei2019, Lei2019b}. How  hyperuniformity changes across the dynamic transition from an absorbing to diffusive state is still an open question and it is the purpose of this letter to address this question in the context of glasses. 
 \\
 In this work,  we focus on a model glass undergoing  yielding when subjected to shear deformation, using computer simulations. Glasses display intriguing properties once submitted to cyclic deformation of various  strain amplitudes, $\gamma_{max}$~\cite{fiocco2013oscillatory}. In particular, it has been shown that there exists a critical amplitude, now identified with yielding $\gamma_{y}$ ~\cite{leishangthem2017yielding}, that marks a dynamic transition  between an absorbing ($\gamma_{max}\le\gamma_y$) and a diffusive ($\gamma_{max}> \gamma_y$) state. Moreover, it has been shown that cyclical shear below yielding can lead to annealing\cite{leishangthem2017yielding,PallabiAnnealing,PallabiPNAS2020}, while above it, the system presents clear evidence of shear banding  related with the rejuvenation~\cite{Parmar2019}. We will show that, indeed, these systems show a good degree of hyperuniformity in the absorbing state but this behaviour changes dramatically across the dynamical transition.

\begin{center}
 \begin{figure}[!htb]
  \centering\includegraphics[width=.85\linewidth]{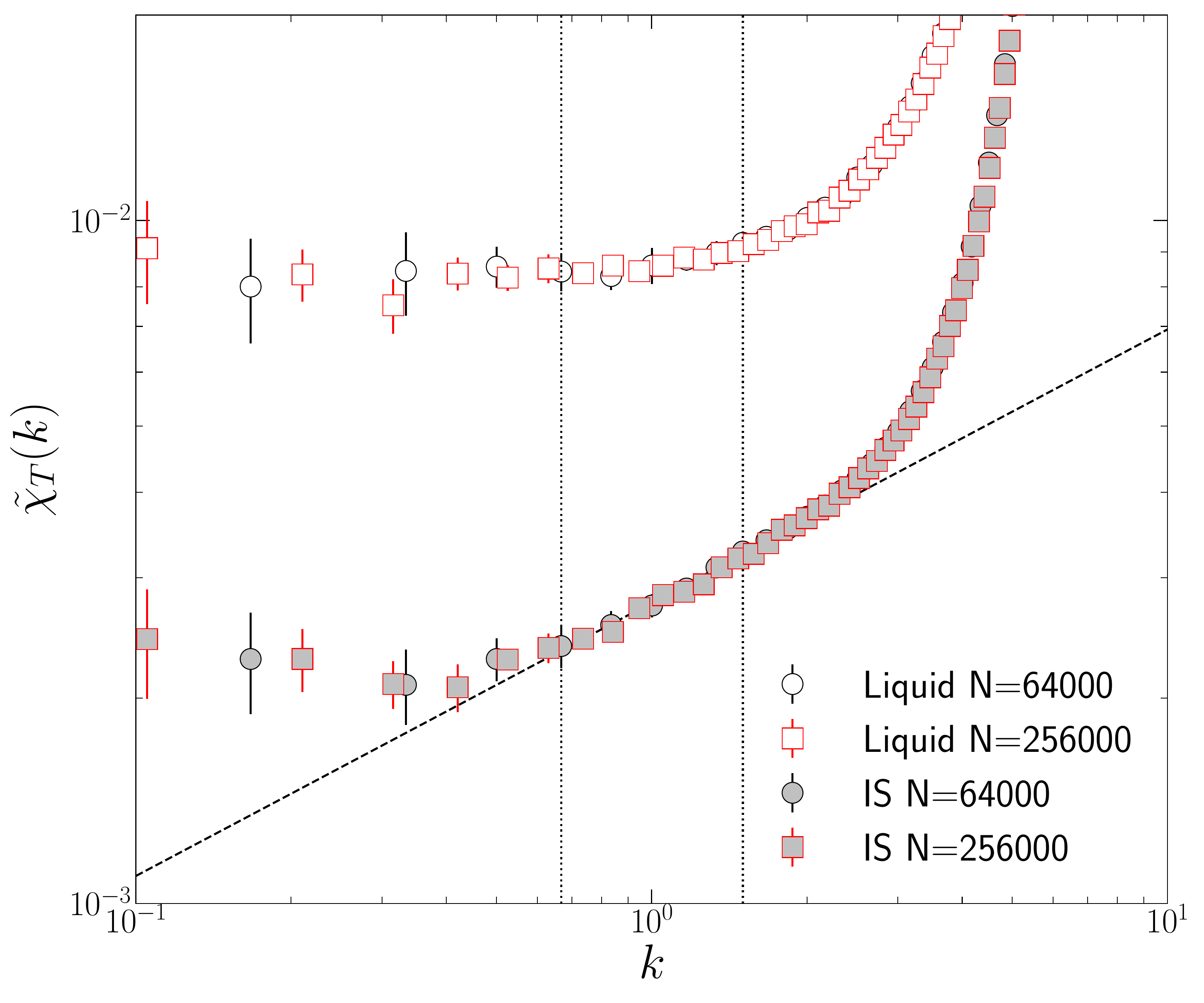}
  \caption{The compressibility $\tilde{\chi}_T(k)$ of the liquid and the corresponding IS at $T=1.0$.  Circles and squares correspond to system sizes $N=64000$ and $N=256000$, respectively. The dashed curve is the power law fit, in the regime marked by vertical lines, with exponent $\alpha \sim 0.40$.}
  \label{Fig:1}
 \end{figure}
\end{center}

Following previous works~\cite{fiocco2013oscillatory,leishangthem2017yielding,Parmar2019}, we explore a model glass, the Kob-Andersen $80:20$ binary Lennard-Jones mixture of of 64000 (or 256000)  particles interacting with the Lennard-Jones potential with a quadratic cut-off (see Supplemental Materials (SM) for further details). 
Under cyclic athermal quasistatic shear (AQS),  the evolution and the steady state of the system are monitored by considering the properties of local energy minimum states or {\it inherent structures} (IS) at zero strain, as a function of the cyclic shear amplitude.

Above a critical strain amplitude { $\gamma_y \simeq 0.070$}, the model exhibits a transition from absorbing to diffusive state \cite{fiocco2013oscillatory}. To examine the existence of a hyperuniform state across yielding, we study density fluctuations {\it via} the wave-vector $k$ dependent compressibility, defined for thermal binary mixtures as 
\begin{align}
 &\rho k_B T \chi_T (\textbf{k}) = \frac{S_{AA}(\textbf{k})S_{BB}(\textbf{k})-S_{AB}(\textbf{k})^2}{c_A^2 S_{BB}(\textbf{k})+c_B^2S_{AA}(\textbf{k})-2c_Ac_B S_{AB}(\textbf{k})}
 \label{chieqn}
\end{align}
where, $c_A=N_A/N$ and $c_B=N_B/N$, and $S_{AA}$ {\it etc.} are partial static structure factors (defined in detail in the SM). 
For hyperuniform systems, it is expected that $\tilde{\chi}_T  (k) \equiv \rho k_B T \chi_T (k) \propto k^{\alpha}$, where $0 < \alpha \leq 1$.
\begin{figure}[!htb]
\centering\includegraphics[width=.9\linewidth]{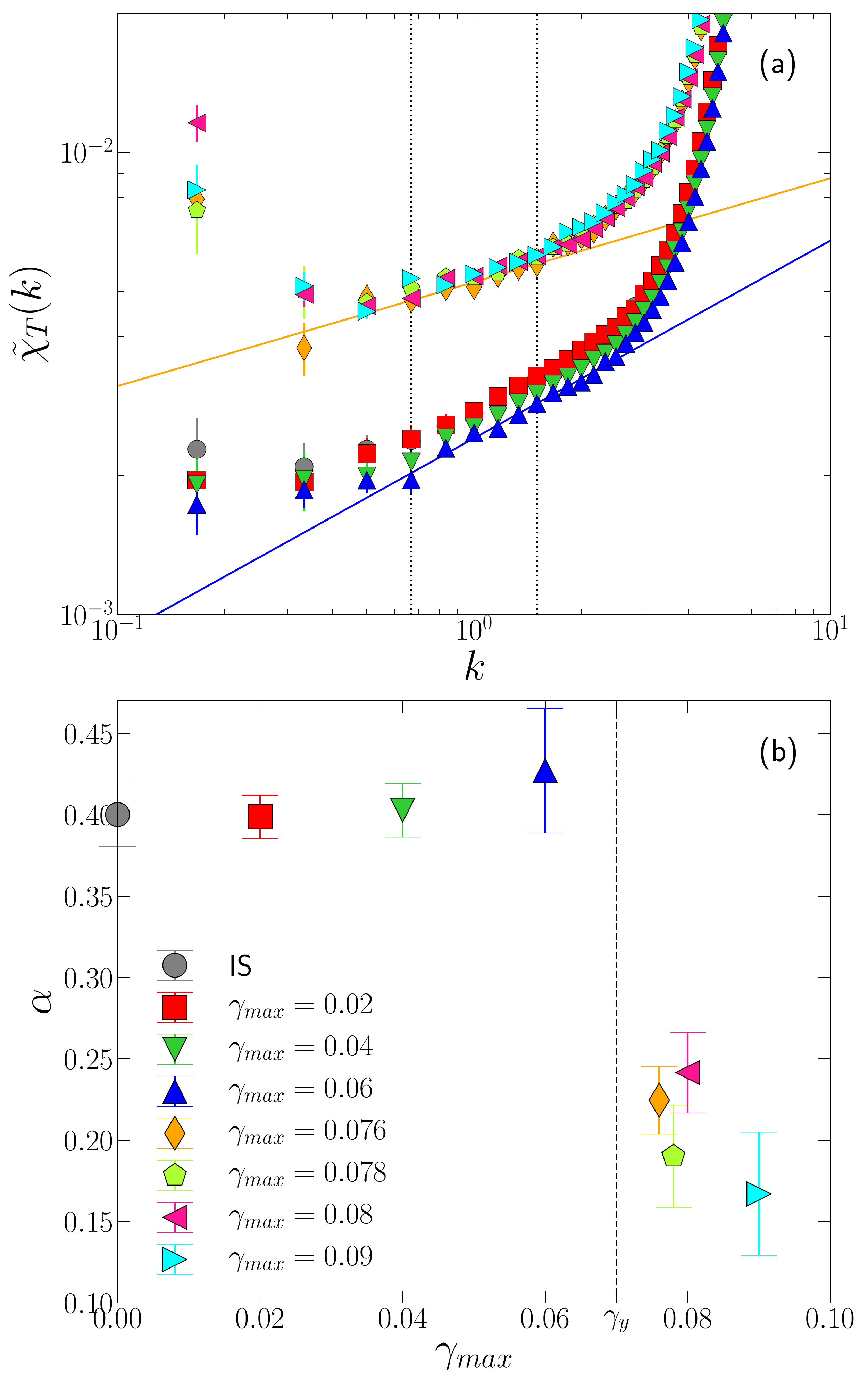}
\caption{(a) Compressibility  $\tilde{\chi}_T(k)$ for the cases of shear amplitudes $\gamma_{max}$ below and above critical yielding amplitude $\gamma_y \simeq 0.07$ are shown with different symbols. Data for the cases above yielding has been shifted upwards for clear visibility. The solid lines (blue and orange) show the power law fit to the compressibility curves. The vertical dotted lines mark the wave vector regime fitted with the power law $k^{\alpha}$.  (b) The  exponent $\alpha$ of power law fit is plotted against shear amplitude $\gamma_{max}$. The vertical dashed line marks the yield amplitude $\gamma_y$.}
\label{Fig:2}
\end{figure}
We first consider the compressibility $\tilde{\chi}_T(k)$ for the  unsheared liquid at $T = 1$ and the corresponding inherent structures, presented in Fig. \ref{Fig:1}. As expected, in the liquid, the $k$-dependent compressibility approaches a constant value as $k \rightarrow 0$. For the IS, Fig.~\ref{Fig:1}  shows that,  within a specific interval of k-vectors,  the compressibility follows hyperuniform behavior. We notice, however, that the $\tilde{\chi}_T  (k\rightarrow 0) $ does not drop to zero but attains a finite value. This behavior has  been observed in several systems that are named  \textit{effectively hyperuniform}~\cite{Klatt2019} and for which the {\it hyperuniformity index} $H$, the ratio of the $k \rightarrow 0$ extrapolation of $\tilde{\chi}_T(k)$ to its maximum value (see SM), is small enough. A value of $H\le 2\times 10^{-3}$ confirms indeed that our system is effectively hyperuniform (see SM). For simplicity, in the rest of the paper, we will continue to refer to effectively hyperuniform simply as hyperuniform. Moreover, in Fig. \ref{Fig:1}, we also confirm that our observations are independent of system size.

%

\begin{center}
 \begin{figure}[]
  \centering\includegraphics[width=.9\linewidth]{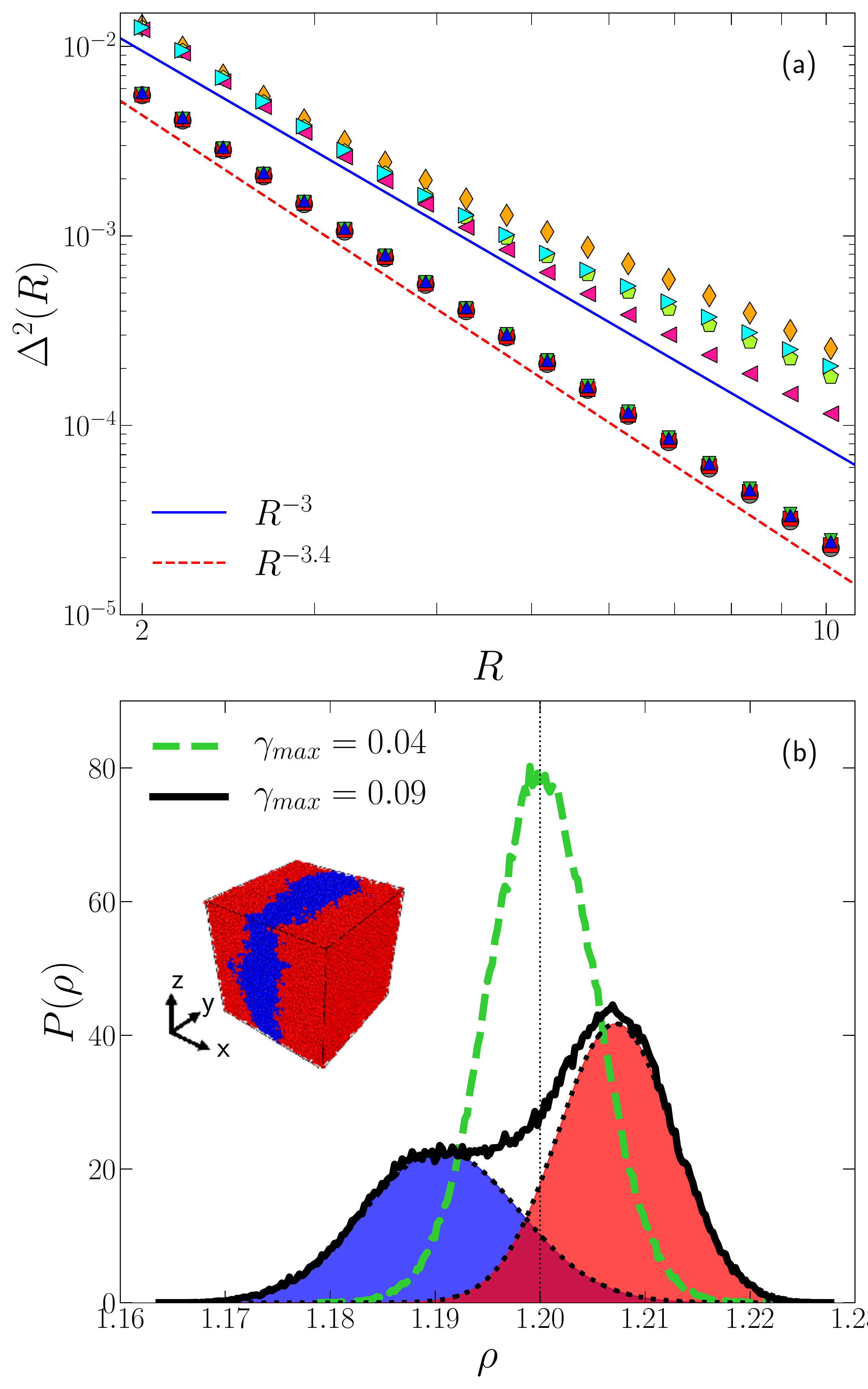}
  \caption{(a) Number density variance as a function of sampled-window radius $R$ for the shear amplitudes $\gamma_{max}$ below and above the yielding amplitude $\gamma_y \sim 0.07$. The symbols used here are same as that of Fig.~\ref{Fig:2}(b). The curves corresponding to amplitudes above yielding have been shifted upwards for better visibility. The variance displays hyperuniformity for $\gamma_{max} < \gamma_y$. High fluctuations are observed for amplitudes above yielding, $\gamma_{max} > \gamma_y$ . (b) Local number density distributions for a window of radius $R=10$ for $\gamma_{max}=0.04$ ($<\gamma_y$) and $\gamma_{max}=0.09$ ($>\gamma_y$). Above yielding, the distribution is bimodal, indicating two distinct high and low density regions in the system. {Inset demonstrates that low density regions are spatially localised. Particles are assigned a colour blue(red) if a window of size $R=10$ around them have a local density $<1.2$($>1.2$).}} 
   \label{Fig:3}
 \end{figure}
\end{center}
\begin{center}
    \begin{figure}[!htb]
          \centering\includegraphics[width=.9\linewidth]{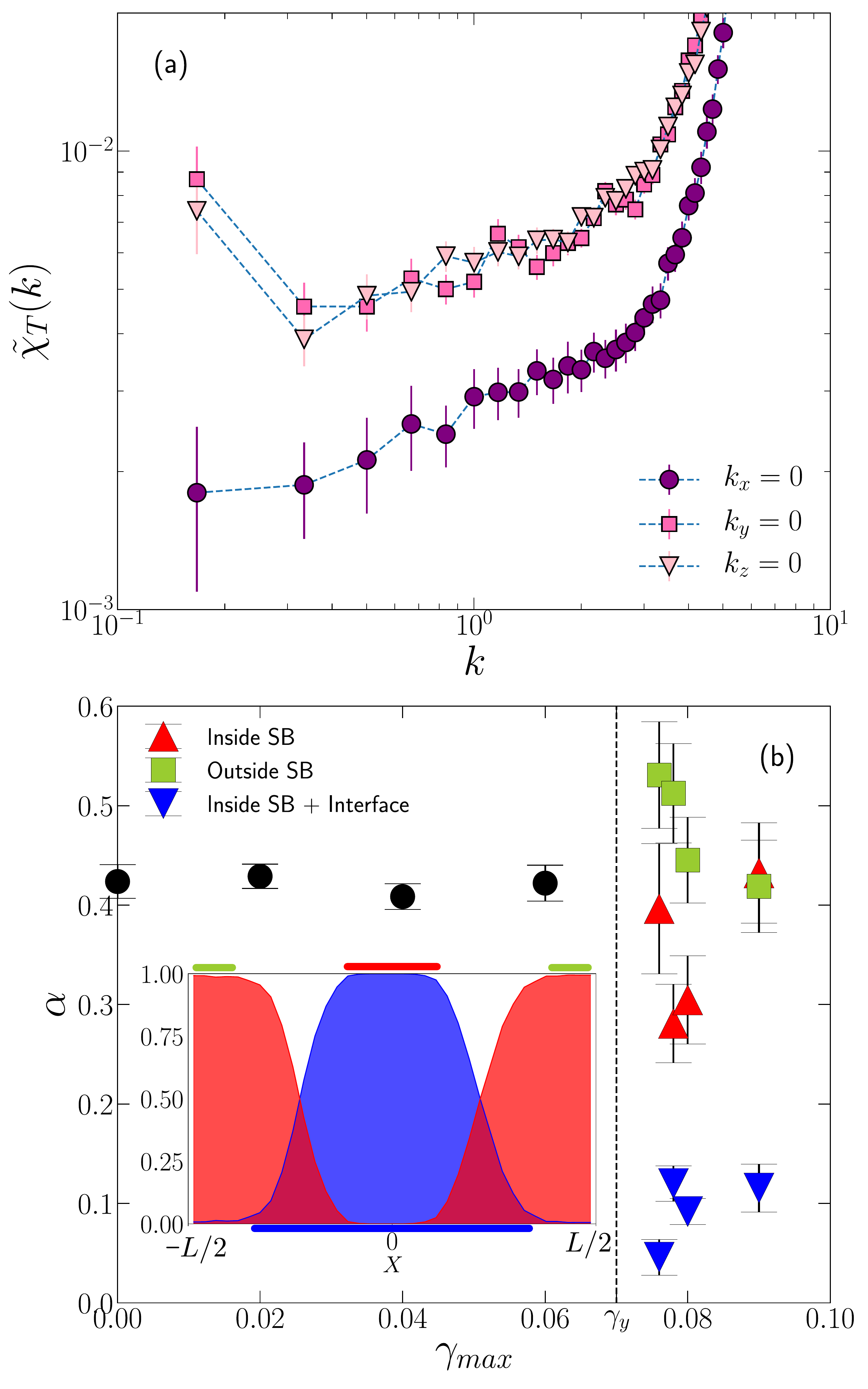}
        \caption{(a) Compressibility $\tilde{\chi}_T(k)$ corresponding to different wave vector orientations for 
     shear amplitude $\gamma_{max}=0.09$ ($>\gamma_y$). Cases (i) $k_x=0$,  (ii) $k_y=0$ and $k_z=0$ correspond to wave vectors being  parallel($k_x=0$) and perpendicular to shear plane. For the perpendicular case, we see an upturn at the lowest wave vectors. (b) Number density variance exponent $\alpha$ (see text for details) for the cases of $\gamma_{max}$ below yielding amplitude $\gamma_y$ are shown as black circles.  Above yielding, for particles inside the shear band, $\alpha$ is shown with red up triangles. For particles outside the shear band $\alpha$ is shown as green squares. When  the interface is included, the exponent drops to low values indicating a lack of hyperuniformity, as shown by the blue down triangles. {Inset: spatial distribution of densities. The red(blue) histograms show the probability that the local density is above(below) $\rho=1.2$. In this way, we clearly identify the regions inside (blue) and outside (red) the shear band. The red(green) bar  on the top of the graph marks the sampling regions fully inside(outside) the  shear band. The blue bar on the bottom marks the region that includes the interface. (\textit{See text for details.})}  }
        \label{Fig:4}
    \end{figure}
\end{center}

Having established that the unsheared energy minimum configurations are  hyperuniform, we turn our attention to systems under  cyclic deformation. The initial IS configurations are periodically deformed following the AQS protocol, for  a range of amplitudes $\gamma_{max}$. Up to $50-500$  cycles are performed to ensure that a steady state has been reached~\cite{fiocco2013oscillatory}. After that, $\tilde{\chi}_T  (k)$  is computed  for different values of $\gamma_{max}$ on different  configurations sampled stroboscopically (see SM for more details). 
The results are shown in Fig.~\ref{Fig:2}(a).  Two different trends emerge, in the absorbing state ($\gamma_{ max}<\gamma_y$), the results for the sheared systems are similar to the unsheared IS - compressibility is suppressed in the absorbing state, as can be clearly observed from the reduced values of H (see SM).  Indeed, from fits to the form $\tilde{\chi}_T  (k) \propto k^{\alpha}$, we obtain values of $\alpha$ around 0.4, as shown in  Fig.~\ref{Fig:2}(b).  
\\
In the diffusive state, $\gamma_{ max}>\gamma_y$ ,  two relevant differences emerge with  respect to the previous case. First, the exponent $\alpha$ changes dramatically as soon as $ \gamma_y$ is crossed, (Fig.~\ref{Fig:2}b). Second, in the long wavelength limit ( $k\rightarrow 0$), a strong upturn of $\tilde{\chi}_T  (k)$ suggest the presence of  large fluctuations. Results in Fig.~\ref{Fig:2} thus suggest an abrupt transition from hyperuniform  to non-hyperuniform  behaviour on crossing the yielding point. This trend is   also confirmed by inspecting the hyperuniformity index $H$ (see SM)~\cite{Torquato2018}.\\
Next, we examine fluctuations in real space. In this case, we introduce the  density variance in real space $\Delta^2(R)$  which is related to the  exponent $\alpha$ defined earlier:
\begin{equation}
\Delta^2(R):= \left<\rho^2(R)\right>-\left<\rho(R)\right>^2 \sim R^{-\left(d+\alpha\right)},
\label{eq:flucreal}
\end{equation}
where $R$ is the radius of the sampling sphere, $\rho$ is the number density within the spherical window of size $R$ and $d$ is the dimensionality. The results of this analysis are presented in Fig.~\ref{Fig:3}(a), where, as with the behaviour of $\tilde{\chi}_T  (k)$, two clear trends emerge. Below yielding, the  exponent $\alpha$ is found to be consistent with those obtained from  compressibility $\tilde{\chi}_T  (k)$ (Fig.~\ref{Fig:2}a).  Above yielding, however, one observes deviations from power law behaviour for large window sizes and the exponent $\alpha$ (if one attempts to estimate it) attains negative values, indicating the presence of strong fluctuations. \\
What is the origin of this puzzling behavior? To answer this question, we investigate in detail the distribution of the local density $\rho$ sampled within a given window size $R=10$. As shown in Fig.~\ref{Fig:3}(b), the local density distribution shows substantially  different behavior below and above yielding. For $\gamma_{ max}<\gamma_y$,  $\rho$ exhibits a unimodal distribution centered around the bulk density. However, above yielding ($\gamma_{ max} > \gamma_y$), the distribution  becomes bimodal and can be  described by the sum of two independent Gaussians. 
Using such a fit, we identify a threshold density $\rho_c$ which we choose to be equal to $\rho =  1.2$. We then study how the low and high density regions are distributed in space. As shown in the inset of Fig.~\ref{Fig:3}(b), this procedure identifies two regions: a low density band (in the middle of the simulation box) sandwiched between a high density region. Indeed, we confirm the correspondence of this low-density sub-volume  to the center of a dynamical shear band discussed in Ref.~\onlinecite{Parmar2019} (see SM). 
\\
The existence of a large interface between two different density regions might be the origin of the low $k$ upturn in the $\tilde{\chi_T}(k)$ observed in Fig.~\ref{Fig:1}  To confirm this expectation, we compute the compressibility,  for the case of $\gamma_{max}=0.09$, restricting the $\mathbf{k}$ vectors to planes parallel  ($k_x=0$) and perpendicular ($k_y=0$ and $k_z=0$) to the shear band plane. The results are displayed in Fig.~\ref{Fig:4}(a). It is evident that, when $k$ is parallel to the shear band, the hyperuniformity features observed below yielding  in Fig.~\ref{Fig:2}(a) are recovered. On the other hand, across the shear bands, the density fluctuations are enhanced by the presence of the interface between  the shear bands. 
\\
At this point, a  question naturally arises, what is the nature of  the fluctuations inside and outside the shear band? To answer this question we  measured  directly $\Delta^2(R)$  inside and outside the shear bands (see SM). 

{To pin down exactly the location of the shear band, we sample the spatial distribution of densities using spherical  windows of size $R=10$ for the case of amplitude $\gamma_{max}=0.09$.  Samples are  classified  using a cut off density $\rho=1.2$. As shown in  Fig.~\ref{Fig:4}b (inset), windows with densities below the cutoff are mostly inside the shear band. 
We first restrict our calculation to regions that exclude completely the interface (as marked on the inset).} The resulting variation of the exponent $\alpha$ at different values of the shear amplitude is presented in Fig.~\ref{Fig:4}(b). {Below yielding, density fluctuations across the system have similar character, and the $\alpha$ values do not depend on the $X$ coordinate, and are the same as the ones obtained in k-space and presented in Fig.~\ref{Fig:2}(b). Above yielding the behaviour is very similar inside and outside the shear band. In both regions,  we recover the same level of hyperuniformity as below yielding. However, if we enlarge the window of sampling in the shear band such that the interface is included (as marked in the inset of Fig.~\ref{Fig:4}(b)), the hyperuniformity is completely lost and similar results as in Fig.~\ref{Fig:2}b are recovered. A similar result is observed for the compressibility as calculated for $k_x = 0$ ({wave-vector along the shear band-plane}) inside and outside the shear band (see SM). We can conclude that the interface between the two regions of different densities, in consequence of shear band formation, is responsible for the disruption of hyperuniformity.  }
In summary, we have investigated the presence of hyperuniformity in a model glass subjected to cyclical deformation. Previous work has established that under cyclic deformation, a sharp boundary may be identified between a pre- and post-yield regime, corresponding to deformation amplitudes {$\gamma_{max} \leq \gamma_y$} and $\gamma_{max} > \gamma_y$ respectively, and these regimes correspond to non-diffusive (absorbing) and diffusive states when one follows the movement of particles from one cycle to the next stroboscopically. With cyclic shear, the glass anneals progressively ~\cite{Parmar2019}, and we show here that it corresponds to an increase in the degree of hyperuniforming compared to the unsheared glasses, which we also show exhibit hyperuniformity. Above yielding, we demonstrate that hyperuniformity is lost as a result of increased density fluctuations associated with the formation of an interface between two regions with different densities.
If we restrict our evaluation of the fluctuation inside and outside this shearband, excluding the interface, the system continues to be hyperuniform in the same manner as the sheared glasses below yielding. Past studies have considered systems which exhibit hyperuniformity homogeneously in space. Here we demonstrate, for the first time, the possibility of coexistence of hyperuniform regions  in a driven system. This observation points to new directions for the study of textured, or modulated, systems with spatially varying degrees of hyperuniformity {with interface}, which may be of great interest to investigate further.

\begin{acknowledgments}
We thank Vinutha H. A. for useful scientific discussions. We gratefully acknowledge IFCPAR/CEFIPRA for support through project no. 5704-1. GF and SM acknowledge the International Centre for Theoretical Sciences (ICTS) for supporting a visit and participation in the program Entropy, Information and Order in Soft Matter ICTS/eiosm2018/08. SS acknowledges support through the JC Bose Fellowship DST (India).
\end{acknowledgments}


\bibliographystyle{apsrev4-1}
\bibliography{references}

\end{document}